\documentclass[a4paper,11pt]{article}
\usepackage{jheppub}
\pdfoutput=1
\usepackage{enumitem}
\usepackage[activate=true]{microtype}
\usepackage{soul,color}
\usepackage{ifthen}

\usepackage[T1]{fontenc}
\usepackage[utf8]{inputenc}

\newboolean{@draft}
\setboolean{@draft}{false}
\ifthenelse{\boolean{@draft}}{  }{  }

\ifthenelse{\boolean{@draft}}{
\usepackage{draftwatermark}
\SetWatermarkText{DRAFT}
\SetWatermarkScale{3.5}
}{}

\title{Update of the Brazilian Participation in the Next-Generation Collider Experiments}

\author[a]{W. L. Ald\'a J\'unior}
\author[f]{G. A. Alves}
\author[a]{K. M. Amarilo}
\author[a]{M. Barroso Ferreira Filho}
\author[e,b]{C. A. Bernardes}
\author[a]{E. M. da Costa}
\author[f, j]{U. de Freitas Carneiro da Gra\c ca}
\author[a]{D. de Jesus Dami\~ao}
\author[a]{S. de Souza Fonseca}
\author[h,f]{L. M. Domingues Mendes}
\author[a]{M. Donadelli}
\author[e,a]{G. Gil da Silveira}
\author[f]{C. Hensel}
\author[o]{C. Jahnke}
\author[a]{H. Malbouisson}
\author[g, i]{J . L. Marin}
\author[n]{D. E. Martins}
\author[f]{A. Massafferri}
\author[a]{C. Mora Herrera}
\author[d]{I. Nasteva}
\author[g]{E. E. Purcino de Souza}
\author[l,m]{F. S. Queiroz}
\author[d]{M. Rangel}
\author[f,k]{P. Rebello Teles}
\author[a]{M. Thiel}
\author[b]{T. R. F. P. Tomei}
\author[a]{A. Vilela Pereira}

\affiliation[a]{
Departamento de Física Nuclear e Altas Energias,\\ Universidade do Estado do Rio de Janeiro (UERJ),\\
Rua São Francisco Xavier, 524, 20550-900, Rio de Janeiro, Brazil}

\affiliation[b]{Universidade Estadual Paulista (Unesp), Núcleo de Computação Científica\\
Rua Dr. Bento Teobaldo Ferraz, 271, 01140-070, Sao Paulo, Brazil}

\affiliation[c]{Instituto de Física, Universidade de São Paulo (USP),\\
Rua do Matão, 1371,  05508-090, São Paulo, Brazil }

\affiliation[d]{Universidade Federal do Rio de Janeiro (UFRJ), Instituto de Física,\\
Caixa Postal 68528, 21941-972 Rio de Janeiro, Brazil}

\affiliation[e]{Instituto de Física, Universidade Federal do Rio Grande do Sul ,\\Av. Bento Gonçalves, 9550,  91501-970, Caixa Postal 15051, Porto Alegre, Brazil}

\affiliation[f]{Centro Brasileiro de Pesquisas Físicas (CBPF),\\
Rua Dr. Xavier Sigaud, 150,  22290-180 Rio de Janeiro, RJ, Brazil}

\affiliation[g]{Escola Politécnica, Universidade Federal da Bahia (UFBA),\\
Rua Aristides Novis, 2  40210-630 Salvador, BA, Brazil}

\affiliation[h]{Laboratório de Instrumentação e Física Experimental de Partículas LIP (LIP),\\
Av. Prof. Gama Pinto 2, 1649-003 Lisboa, Portugal}

\affiliation[i]{Coordenação de Eletrônica (CELET), Instituo Federal da Bahia (IFBA), \\ Av. Sérgio Vieira de Melo, 3150 - Zabelê, Vitória da Conquista - BA, 45078-300  }

\affiliation[j]{Coordenação de Engenharia Eletrônica (CCGELT), \\ Centro Federal de Educação Tecnológica Celso Suckow da Fonseca (CEFET/RJ), \\ Av. Maracanã, 229,  20271-110, Rio de Janeiro, Brazil  }

\affiliation[k]{CERN, EP Department 1211 Geneva, Switzerland}

\affiliation[l]{Departamento de F\'{\i}sica, Universidade Federal do Rio Grande do Norte,\\ 59078-970, Natal, RN, Brasil}

\affiliation[m]{International Institute of Physics,  Universidade Federal do Rio Grande do Norte, \\Campus  Universit\'ario,  Lagoa  Nova,  Natal-RN  59078-970,  Brazil}

\affiliation[n]{Nuclear Physics Institute, im. Henryka Niewodnicza\'{n}skiego, \\
Polskiej Akademii Nauk, ul. Radzikowskiego 152,
31-342 Krakow, Poland}

\affiliation[o]{Universidade Estadual de Campinas (UNICAMP), Instituto de F\'isica Gelb Wataghin, R. S\'ergio Buarque de Holanda, 777 - Cidade Universit\'aria, Campinas, SP, Brazil.}

\emailAdd{gustavo.silveira@cern.ch}


\abstract{
This proposal outlines the future plans of the Brazilian High-Energy Physics (HEP) community for upcoming collider experiments. With the construction of new particle colliders on the horizon and the ongoing operation of the High-Luminosity LHC, several research groups in Brazil have put forward technical proposals, covering both hardware and software contributions, as part of the Brazilian contribution to the global effort. The primary goal remains to foster a unified effort within the Brazilian HEP community, optimizing resources and expertise to deliver a high-impact contribution to the international HEP community.

Paper submitted to the Latin American Strategy Forum for Research Infrastructure
}

\begin{document} 
\maketitle

\section{Scientific Context}
\label{sec:scicont}

The advent of the Large Hadron Collider (LHC) as a proton-proton ($pp$) collider has provided large amounts of data at higher energies than its predecessors. Several topics are investigated with the LHC data collected since 2022 with the start of the Run-3, the third data-taking period, with great focus in investigating the determination of the parameters of the Standard Model (SM) at high precision \cite{CMS:2024gzs,ATLAS:2024wla,Martelli:2024uvh} and searches for New Physics \cite{Knapen:2022afb,ATLAS:2024fdw} based on proposals on the literature. The LHC operation timeline aims to collected data in Run-3 with planned operation until 2025. In 2023 there was a heavy-ion run of Pb-Pb collisions, where an unprecedented amount of data has been collected by the LHC experiments, leading to the opportunity of investigating the quark gluon plasma (QGP) and hydrodynamics at high precision. As a result, we expect an improvement in statistical and systematic uncertainties lead to better theoretical description of Monte Carlo (MC) simulations and higher-order corrections. 


The expectation with the upcoming period with higher luminosity, called Phase 2 or \textit{High Luminosity LHC} (HL-LHC) \cite{ZurbanoFernandez:2020cco}, LHC program will provide even larger sets of data to improve the searches for New Physics. The operation starting in 2029 will collect nearly 10 times more data than Phase-1, reaching 3000/fb by the 2040s. Improving the searches for New Physics is the main goal of this operation at higher luminosity \cite{ellis-esppu-2019}. Hence, the LHC experiments have been planning their upgrades to since many years \cite{cernLongTerm} and will start the installation and commissioning during the Long Shutdown 3 to start in 2026.


Looking beyond the LHC, there are new projects for future accelerators to be built all over the world:

\begin{itemize}

\item Circular Electron Positron Collider (CEPC): a 100~km $e^+e^-$ collider to be built in China to starting taking data by 2040, aiming to be a Higgs factory \cite{CEPCStudyGroup:2023quu};

\item International Linear Collider (ILC): construction planned to happen in Japan, a 20~km linear $e^+e^-$ collider to achieve colliding energies of 1~TeV \cite{Behnke:2013xla};

\item Compact Linear Collider (CLIC): similar to ILC, a linear collider of 50~km at CERN to achieve energies of 1.5~TeV and 3.0~TeV in two operation phases \cite{Charles:2018vfv};

\item Future Circular Collider (FCC): a 90~km circular collider at CERN comprising collisions of $e^+e^-$, $eh$, $hh$ up to 100~TeV and to start its first phase by 2048 and $hh$ by 2070 \cite{FCC:midterm};

\item Electron-Ion Collider (EIC): project already approved by DoE and to start construction at Brookhaven National Laboratory, it will collide electron and ions at energies of \cite{Accardi:2012qut}.

\end{itemize}

Among these projects, only EIC is meant to start taking data by 2030, with C and FCC planning their start at 2040 and 2049, respectively.

\section{Physics motivation and objectives}


\subsection{Central Exclusive Production}

The search for evidence of New Physics has been an intense topic of investigations in the HEP community. In particular, the coupling of new particles to the electroweak sector poses an opportunity for searches of new particles at the LHC. A fraction of the hadronic collisions correspond to the interaction via photons and pomerons, leading to the so-called \textit{Central Exclusive Production} (CEP). The experimental signature of such an event is the presence of intact protons in the final state and low hadronic activity due to the exchange of colorless particles, resulting in a gap in pseudorapidity in the detector between the protons and the produced central system. Although usual technologies are employed to observe the particles in the central detector, near-beam detectors have to be used to observe the intact protons (so-called \textit{proton tagging}), which are scattered in very small angles ($\sim$mrad). The key physical goal of studying  events is the search for New Physics in the coupling of new particles to the electroweak sector and improve the modeling of underlying events in hadronic collisions. In addition, there are opportunities for studying single tagging in $p$A collisions at the LHC, especially with light ions.


\subsection{Heavy Ion Physics challenges}

The heavy ion (HI) physics program in Run~3 and 4 of LHC will receive great benefits through the increase of collision luminosities, detectors upgrades, and new data analysis techniques. For example, high-statistics samples of high-purity heavy-flavor hadrons with very low transverse momentum ($p_{\mathrm{T}}$), high-efficiency charged particle reconstruction (tracking) down to low-$p_{\mathrm{T}}$ (few hundreds MeV) up to almost 1~TeV, and enriched samples of moderate-$p_{\mathrm{T}}$ jets ($p_{\mathrm{T}}\approx30\mbox{--}100~\mathrm{GeV}$) is planned to be available. The four main collaborations at the LHC are strongly involved in this endeavor~\cite{CMS:2024krd, ALICE:2022wpn, Aad:1129811, LHCb:2008vvz}, which is vital 
for the understanding of two fundamental theories of the SM of particle physics, Quantum Chromodynamics (QCD) and Quantum Electrodynamics (QED). In special, this sector is reaching a precision era, willing to improve our understanding of the thermodynamic and transport properties of the QGP, a deconfined state of quarks and gluons created in heavy ion collisions~\cite{CMS:2024krd, ALICE:2022wpn}. At the same time, new QGP probes are being discovered (e.g.,~\cite{Andrews:2018jcm,Andres:2022ovj,CMS:2024sgx}). Some of the main topics expected to receive important inputs from this program are~\cite{CMS:2024krd, ALICE:2022wpn}:

\begin{enumerate}

    \item Search for the onset of parton saturation by performing measurements constraining gluon densities in the Pb nucleus at very low Bjorken-$x$ (down to $10^{-6}$) with unprecedented precision;
    
    \item Search for the onset of collective effects in small colliding systems, characterization of the nuclear parton distribution functions (nPDFs), and modeling of ultrahigh-energy (cosmic ray) phenomena; In special, the pilot runs of oxygen-oxygen (OO) and proton-oxygen ($p$O) will provide inputs on these three topics;
    
    \item Studies using Ultra-peripheral HI Collisions (UPCs) together with high-precision QED calculations, will give important inputs for the search for new physics in the low-mass range, and understanding of heavy-flavor production, among others. In special, the study of the final state of two jets emitted back-to-back in UPCs, with virtually zero backgrounds, provides a unique opportunity to probe fundamental topics in physics: extraction of the strong coupling constant, refinement of charm jet flavor labeling algorithms, and improvements in the accuracy of parton shower predictions;
    
    \item The hard probe sector, providing information about the strongly coupled QGP, will be boosted by the novel tools to investigate the jet quenching phenomenon, together with high statistics high-$p_{\mathrm{T}}$ hadrons, fully reconstructed jets, heavy quarkonia and open heavy-flavor particles;

      \item Measurement of exotic mesons in new kinematic ranges not acessible before. For example, the measurement of X(3872) is possible by ALICE detector using Run-3 data, which is complementary to previous measurements. This provides information and constraints to model calculations to distinguish between a compact tetraquark state from a dilute molecular state.
    
    \item Study of the 3D nucleus/proton structure, which can be accessed by Multiple Parton Interactions (MPI) observables~\cite{dEnterria:2017yhd}. In addition to the structure of the nucleus/proton, the study of MPI also provides inputs in the search for new physics, especially in the characterization of rare final-state backgrounds.
    
\end{enumerate}

In addition, this physics program will complement the studies from RHIC~\cite{PAC_BNL} and the upcoming EIC~\cite{Accardi:2012qut}.

\subsection{Higgs boson pair production and Higgs boson self-coupling}

Following the discovery of the Higgs boson ($H$) in 2012~\cite{HIGG-2012-27,CMS-HIG-12-028}, the properties of this new boson have been widely investigated. To date, all measurements of its quantum numbers and couplings are in excellent agreement with the SM~\cite{ATLAS:2015zhl,ATLAS:2020ior,CMS:2021nnc,ATLAS:2023dnm,CMS:2022ley,ATLAS:2022vkf,CMS:2022dwd}. However, the properties of interactions involving several Higgs bosons still need to be verified, being the observation of the production of Higgs boson pairs $(HH)$  the next milestone in the physics program of the LHC, due to their role in cosmological theories involving, for example, vacuum stability and inflation~\cite{Bass:2021acr}.  Of all the self-interaction terms of the Higgs boson, the trilinear term is the only one in the reach of the HL-LHC. Deviations of the trilinear self-coupling  from the SM value  play a fundamental role in many of the current theories of BSM physics, chief among them Electroweak Baryogenesis~\cite{Morrissey:2012db}.
The most sensitive test of Higgs boson self-interactions comes from processes of $HH$ production via gluon-gluon fusion (ggF) and vector-boson fusion (VBF). Measuring the cross-section of  these processes offers a direct probe of the values of these couplings, through their scale factors with respect to the SM predictions: $\kappa_{\lambda} = \lambda_{HHH}/\lambda^{\rm{SM}}_{HHH}$, affecting both ggF and VBF production, and $\kappa_{V} = g_{HVV}/g^{\rm{SM}}_{HVV}$ and $\kappa_{2V} = g_{HHVV}/ g^{\rm{SM}}_{HHVV}$, which only impact VBF production.  


The small SM production cross-sections of these processes implies that the main
experimental signatures will be those where at least one of the two Higgs bosons decays into a final state with a large branching ratio, i.e. $H \rightarrow b\bar{b}$. Among all the possible channels, the most sensitive signatures are $HH \rightarrow b\bar{b}b\bar{b}$, $HH \rightarrow b\bar{b}\tau^{+}\tau^{-}$ and $HH \rightarrow b\bar{b} \gamma \gamma$, with branching ratios of 33.9\%, 7.3\% and 0.26\% respectively. Ne\-vertheless,  because of its small cross-section, approximately 1000 times smaller than that of single Higgs, $HH$ production has long been considered out of the LHC reach. However, owing to the spectacular improvement of $HH$ analyses over the past decade, $HH$ production is now being investigated in several production and decay modes, setting stringent constraints on new physics and rapidly approaching the SM prediction sensitivity with recent Run 2 results from ATLAS and CMS collaborations, using about $140$~fb$^{-1}$ of $pp$ collision data at $\sqrt{s} = 13$~TeV. At 95\% confidence level (CL), ATLAS results on the observed (expected) upper limit on the $HH$ production rate is 2.9 (2.4) times de SM prediction~\cite{ATLAS:2024ish}, whereas CMS obtained 3.4 (2.5) times de SM prediction~\cite{CMS:2022dwd} with Run 2 data. Due to the different experimental advantages of each decay channel, the analyses are complementary to target BSM physics effects in $HH$ production. 

The current LHC data taking period - Run 3, with $pp$ collisions at $\sqrt{s}=13.6$~TeV, brings a series of improvements for the $HH$ search beyond the increase in the production cross-section ($\sigma_{gg\rm{F}+VBF}$) that scales with the center-of-mass collision energy, such as: more refined analysis techniques;  major developments to $b$-tagging algorithms and new dedicated trigger chains for $HH \rightarrow b\bar{b}b\bar{b}$ and $HH \rightarrow b\bar{b}\tau^{+}\tau^{-}$ channels, benefiting the increase in signal acceptance. Thus, the SM expectation for $HH$ production in ATLAS is within reach for Run 3,  assuming the same pace of analysis improvements from Run 2 (a factor of 1.7) and/or combination with CMS.

\begin{figure}[t]
    \centering
    \includegraphics[width=0.47\linewidth]{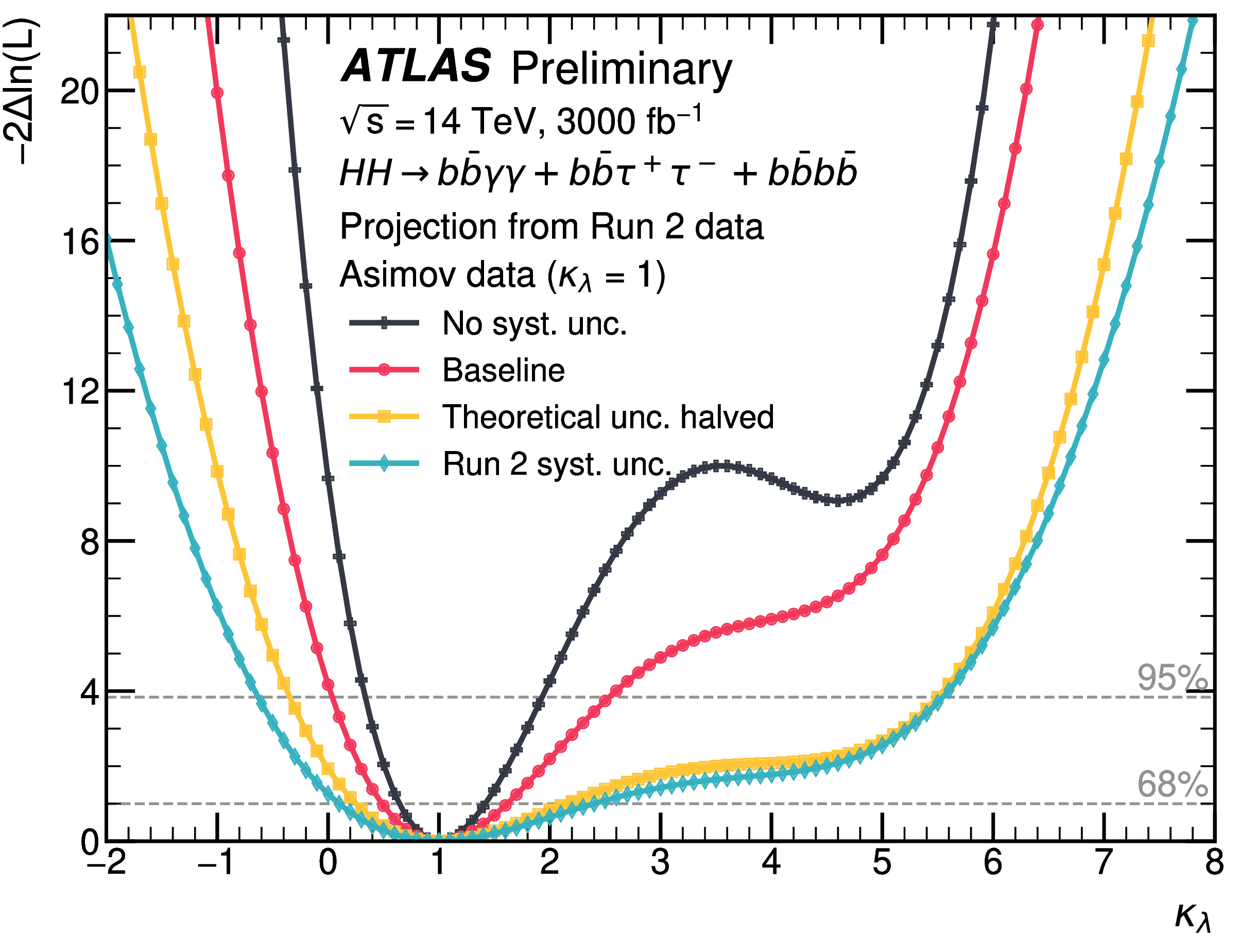}
    \includegraphics[width=0.47\linewidth]{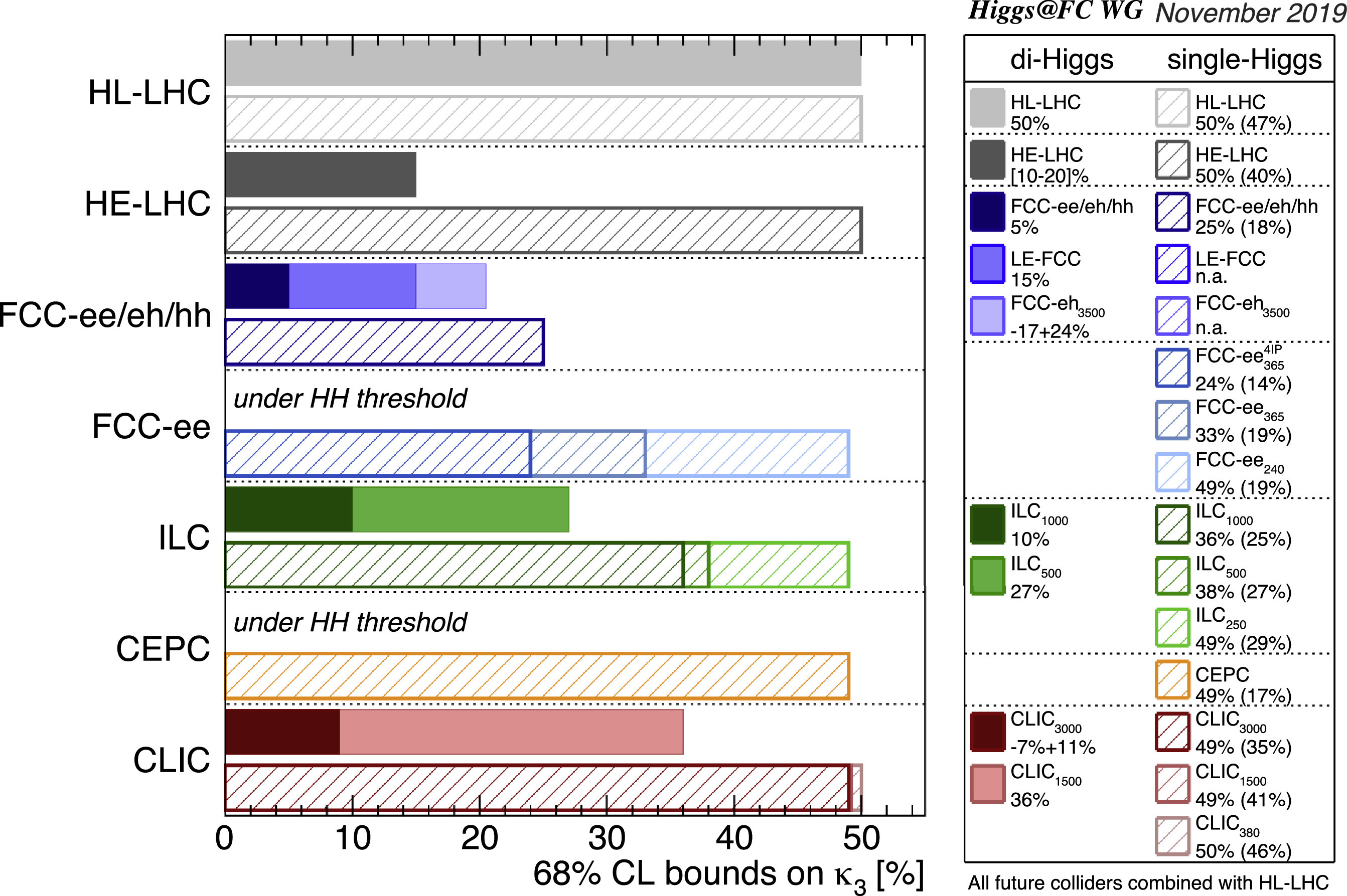}
    \caption{(Left) Projected negative log-profile-likelihood as a function of $\kappa_{\lambda}$ evaluated on an Asimov dataset constructed under the SM hypothesis of $\kappa_{\lambda}=1$, combining $HH \rightarrow b\bar{b}b\bar{b}$ and $HH \rightarrow b\bar{b}\tau^{+}\tau^{-}$ and $HH \rightarrow b\bar{b}\gamma \gamma$ channels  at 3000 fb$^{-1}$ and $\sqrt{s}$ = 14 TeV, assuming the four uncertainty scenarios described in Ref.~\cite{ATL-PHYS-PUB-2022-053}. The intersections of the dashed horizontal lines with the profile likelihood curve define the 68\% and 95\% confidence intervals, respectively. (Right) Uncertainties on the Higgs self-coupling projected for the High-Luminosity LHC and for other future colliders, at various stages, by the ECFA Higgs$@$Future Colliders working group~\cite{deBlas:2019rxi}. The results are presented as uncertainties on $\kappa_{3} = \kappa_{\lambda}$~\cite{MICCO2020100045}.}
    \label{neg-log-prof-likelihood-hh-future}
\end{figure}

Furthermore, the HL-LHC will collide protons at 14 TeV, which constitutes an increase of $\sigma_{gg\rm{F}+VBF}$ that combined with a 20-fold increase in luminosity with respect to the Run 2 dataset (expected to produce an integrated luminosity of 3000 fb$^{-1}$ per interaction point), yields numerous $HH$ events that make the HL-LHC a great facility   to discover and measure the Higgs self-coupling.  Previous combined projection studies of ATLAS~\cite{ATL-PHYS-PUB-2018-053} and CMS~\cite{CMS-PAS-FTR-18-019}, showed that a discovery significance of 4.0$\sigma$ could be achieved, and when systematic uncertainties were neglected, statistical significance was shown to increase to 4.5$\sigma$. 
More recently, ATLAS updated the HL-LHC projection studies with various systematic uncertainty scenarios, and from a statistical combination of $HH \rightarrow b\bar{b}b\bar{b}$ and $HH \rightarrow b\bar{b}\tau^{+}\tau^{-}$ and $HH \rightarrow b\bar{b}\gamma \gamma$ channels, a discovery significance of 4.9$\sigma$ is expected, when systematic uncertainties are neglected~\cite{ATL-PHYS-PUB-2022-053}.  In the same uncertainty scenario, the total 68\% CL interval is expected to be $0.7 < \kappa_{\lambda} < 1.4$ as shown in Figure~\ref{neg-log-prof-likelihood-hh-future} (left). In the next years, further improvements of the offline reconstruction and analyses techniques will help to optimally exploit the upgraded ATLAS detector towards the discovery of non-resonant $HH$ production at the HL-LHC.

The goal for future machines beyond the HL-LHC should be to probe the Higgs potential quantitatively. Such a level of precision is achievable through the measurement of $e^{+}e^{-}$ production at lepton machines at energies above 500 GeV and at hadron machines (FCC-hh). Given likely improvements to ATLAS and CMS (HL-)LHC analyses, single-Higgs precision measurements with a $e^{+}e^{-}$ machine are unlikely to add anything to measurements of $\kappa_{\lambda}$ (unless a high energy linear collider), as shown in Figure~\ref{neg-log-prof-likelihood-hh-future} (right)~\cite{MICCO2020100045}. That said, HL-LHC constraints on $\kappa_{\lambda}$ will have unmatched precision for an extremely long time.

\section{Scenario changes since last report}

With the goal of establishing an organized community in Brazil, the National Network for High Energy Physics (Rede Nacional de Física de Altas Energias - RENAFAE) has submitted a proposal to the initiative of National Institutes promoted by CNPq, aiming the consolidation of hardware development, mobility, promoting the training of human resources, and dissemination of High-Energy Physics (HEP) in Brazil. The so-called {\texttt{INCT CERN Brasil}} is a result of the synergy of the HEP community contributing to the CERN Collaborations and is organized in such a way that the technologies employed in the experiments have a central role instead of individual Collaborations. Hence, INCT CERN Brasil has the goal of establishing concrete advances in R\&D of detectors within 5 years and consolidating the Brazilian participation in the CERN experiments. Hence, the CERN Collaboration in Brazil have been restructured in order to have faster communication and bring together different research groups to increase their impact.

As of March/2024, Brazil has become an Associated Member of CERN, allowing broader representation and opening opportunities with the Brazilian industry. This association is expected to trigger new collaborations within the HEP community in Brazil, especially in the R\&D of new detectors for the HL-LHC using common technologies. In terms of funding sources, discussions have started in order to secure resources for mobility of researchers and predictability of the calls by the Funding Agencies.

Given the green light given by DoE for the construction of EIC at BNL, Brazilian groups have expressed interest in joining the Collaborations to develop studies on Nuclear Physics. Meeting in the past years have attempted to foster new teams to contribute with hardware and software.

\section{Current status and expected challenges}

In view of the launch of the INCT CERN-Brazil project, it is natural for the Brazilian community to set their sights on the next steps.
The most immediate challenge is the fact that the INCT project has a limited duration (5 years);
it is imperative that the community identifies evergreen funding lines that are more adequate to the timescales of HEP projects.
This is particularly critical now that the discussion on the next-generation collider is starting to converge.

After the INCT project, the next natural milestone would be the creation of a National Laboratory -- a permanent organization that would be dedicated to fostering scientific research and technological development in the context of HEP in Brazil. 
The creation of such a structure is, however, beyond the scope of any single project and/or funding line available in the country.

In that regard, a possible way forward would be to follow the steps of the
Brazilian Center for Research in Energy and Materials 
(Centro Nacional de Pesquisa em Energia e Materiais -- CNPEM).
It was established by Brazilian law in 1998 ``to promote and contribute to scientific and technological development and innovation in Brazil, as well as for training through its national laboratories, research units, and teaching institutions''~\cite{site-CNPEM}.
An equivalent effort for a posited National Laboratory for HEP would need a similar legal framework, supported by a series of interlinked projects that would furnish the necessary scaffolding for the initial laboratory setup.
Similar structures from other countries include the National Institute for Nuclear Physics (Istituto Nazionale di Fisica Nucleare -- INFN) in Italy and the National Scientific User Facilities under the United States Department of Energy.


\section{Methodology}

\subsection{Contributions for hardware development}



\subsubsection{Ultra-fast semiconductor detectors for CMS-PPS}

Semiconductor particle detectors are among the most advanced detector technologies in HEP and have been employed extensively due to their precision and favorable characteristics. 
Due to their very high spatial resolution, achieved by applying sensor segmentation into fine strips or pixels, they are almost always the detectors of choice in tracking and primary and secondary vertex reconstruction. 
Their mechanical stability and possibilities for different geometries and arrangements allow for ease of handling and operation both in normal conditions and in vacuum at low temperature. 
Their compact design also features high time resolution with short signal collection times, as well as the low material budget necessary to minimize the disturbance on particle trajectories and energy loss. 
An excellent energy resolution and linearity of detector response is also achieved.
Semiconductor sensors can be designed as charged particle and photon detectors; they are also used as active detector layers in calorimeter systems.


The main current technological challenges faced in semiconductor detector development are pushing the frontier of time resolution to develop ultra-fast sensors and their corresponding electronics on one hand, and pushing for ever lower detection thresholds to achieve quantum detectors on the other. In addition to these challenges, their use in increasingly more intense radiation environments motivates the search for radiation hard devices and improvements in cooling technologies. 
The new requirements and opportunities are leading to new technologies, which need to be evaluated and optimized in terms of radiation hardness, high granularity and 4D reconstruction (tracking+timing) capabilities~\cite{Detector:2784893}.

\pagebreak
\begin{itemize}
\item Silicon detectors
\end{itemize}

The challenges presented by the LHC experiments have motivated an intense R\&D program on semiconductor sensors for radiation detection towards smaller, faster, and more radiation hard devices for tracking, timing, and energy measurements.

State of the art in the development of Ultra-Fast Silicon Detectors (UFSD), devices that are capable of timing resolutions of few picoseconds, came with the introduction of Low Gain Avalanche Diodes (LGAD)~\cite{Sadrozinski:2013nja,Pellegrini:2014lki,Cartiglia:2015iua,Sola:2017zty,Cartiglia:2019fkq,Sola:2018laf}, a novel structure pioneered by the RD50 Collaboration~\cite{RD50} that incorporates a very thin intrinsic charge multiplication layer in order to fulfill the performance requirements for the HL-LHC experiments and beyond.

The LGADs are fabricated in a high resistivity p-type silicon substrate, where a very shallow n++ implantation is used on top of a deeper p+ layer of few tens of microns. The purpose of this geometry is to produce an intense electric field ($\approx2\,\textrm{kV}/\textrm{mm}$) once a voltage is applied across this junction, hence generating charge multiplication by the acceleration of the primary charges produced by the radiation interacting with the semiconductor material.
Contrary to other fast devices like avalanche photodiodes (APD), this geometry does not enter in self-sustained avalanche regime as the gain is kept between 20 to 50 times (hence the low gain attribute).
The current created by the moving charges (holes) are collected by the electrodes that connect the substrate and the implantation layer, leading to a very fast and high amplitude pulse as a result of the small sensor thickness and the intrinsic charge multiplication.
Low signal jitter, necessary to reach timing resolutions of the order of few picoseconds can be achieved by using thin sensors (tens of microns), leading to a fast signal rise time while the charge gain remains independent of the sensor thickness.
Moreover, the very narrow pulses produced by LGADs ($\approx 1\,\textrm{ns}$ width) allow the operation of these sensors in extremely high rate environments.

The LGAD can be produced either as individual sensors (pixels) or in modules segmented as arrays of pixels. Arrays of pixels of $1.3 \times 1.3~\textrm{mm}^2$ are foreseen to be used by the ATLAS~\cite{2137106} and CMS~\cite{Butler:2019rpu} Phase-II upgrade timing detectors.
The CMS Precision Proton Spectrometer~(PPS) \cite{CMS:2014sdw} upgrade project~\cite{CMS:2021ncv} currently plans the use of a variation of the LGAD technology, with smaller no-gain inter-pad regions~\cite{Paternoster:2020qkd}, and smaller pad sizes in one direction in order to reduce the occupancy in such a near-beam detector. 


\begin{itemize}
\item Synthetic diamonds detectors
\end{itemize}

The use of diamond sensors as ultra-fast detectors of high-energy protons has been used in the CMS and TOTEM Collaborations since 2016 in PPS. Such an apparatus provide the access to exclusive events at the LHC, turning it into a photon collider. Similar physics have been studied in previous colliders like HERA at DESY, but the amount of data collected at the LHC are allowing more detailed studies about the nature of  of pairs, such that dileptons, dibosons, and also improving techniques for searching new particles like missing $\slash\!\!\!\!E_T$ \cite{Bellora:2022kzz}.

Synthetic diamond sensors~\cite{RD42,Borchelt:1994mt,RD42:1999ofw,RD42:2006mib,RD42:2022tdb} are wide band-gap (WBG) semiconductors featuring reduced leakage current and hence low noise, albeit with smaller generated signals. Due to the low noise levels, they can be operated at higher temperatures. Diamond sensors are radiation tolerant. They feature high breakdown levels and can be operated at high electric fields. The diamond sensors are produced as poly- and single-crystalline by chemical vapor deposition~(CVD), followed by a metallization step allowing for different electrode geometries.
They can also be produced with 3D electrode geometries~\cite{Bachmair:2015iba}.
Diamond sensors are capable of high-resolution spatial and timing measurements, with large radiation tolerance \cite{Bossini:2023uwa}. The CMS team at UFRGS is working in developing the metallization process in Brazil with SC diamonds acquired from a Brazilian company. Hence, the in-house production can secure the preparation and characterization of new sensors to PPS in Run-3 and HL-LHC.

\begin{itemize}
\item 3D Pixel Detectors
\end{itemize}

Novel 3D silicon pixel sensors~\cite{Parker:1996dx,DaVia:2009zzb,DaVia:2012ay,Alagoz:2012dt,Pellegrini:2013nq,Bubna:2014ria,Lange:2016jbm,VazquezFurelos:2016dks,Lange:2018paz} have been used in the upgrade of the ATLAS and CMS inner tracker systems, and in the CMS PPS.
They are characterized by columnar electrodes for the signal readout, perpendicular to the sensor surface, in a three-dimensional configuration (hence 3D).

This topology optimizes the charge carrier collection. 3D pixel sensors achieve excellent spatial resolution due to small pixel sizes, are radiation-tolerant --- sustaining up to ${\cal O}(10^{16})\,\text{n}_{\text{eq}}/\text{cm}^2$ -- and feature a small inactive edge.
Due to their fast charge collection time, in addition to high spatial resolution and large radiation tolerance, 3D sensors are currently considered as detectors capable of 4D reconstruction, featuring high timing resolution~\cite{Kramberger:2019ygz,Lai:2020gqq}.
3D sensors are bump-bonded to readout chips (ROCs) specialized to their application. 

\subsubsection{HGTD for ATLAS Phase-II Upgrade}


The ATLAS experiment is preparing the construction of a new high segmented detector subsystem able to provide timing information to the reconstructed tracks with a resolution of 30 ps and covering the region between $2.4<|\eta|<4.0$. This new detector, called High Granularity Timing Detector (HGTD) is conceived as two discs (one each side of the collision point) of 2~m diameter holding the sensors, front end electronics, cooling and mechanical support symmetrically placed from 3~m of the collision point, and to be fit into a space between the barrel and endcap of ATLAS detector~\cite{CERN-LHCC-2020-007}. Given the very limited space, this disc can be at most 12~cm thick, including the neutron moderator for protection of the detectors and electronics upstream. Each of the two ATLAS HGTD discs will require $1.8$ million sensors (about 3.1~m$^{2}$ of silicon) assembled as 4 detector planes, each plane subdivided in 3 rings: inner, intermediary and outer ring. For the sensors, the only viable solution is to use semiconductor devices, since they can be fabricated with high segmentation and thin enough to allow the installation of several detector planes without a substantial increase of the material in front of the calorimeters. The use of ultra-fast timing detectors is a solution being sought not only by ATLAS but also by CMS and LHCb collaborations as one of the most promising technologies to deal with the increased pileup in HL-LHC. It is also one of the critical components being explored for the detectors in the colliders of the post-LHC era. 

The signal from each LGAD channel in the HGTD will be proceeded by front-end electronics composed of a voltage amplifier, a discriminator and 2 TDCs. One TDC will be responsible for measuring the timing between the LGAD signal and the LHC clock phase (thus providing the timing to be associated with the track) and the other the amplitude of the pulse via Time over Threshold (ToT). All this information will be serialized and assembled with data from other channels for transmission to the back-end electronics. This circuitry will be implemented in an ASIC (called ALTIROC), and each ASIC bump bonded to 2 arrays of $15 \times 15$ detectors. The sensors and ASICS will be assembled in modules that will be disposed across the disc area in four detector planes. The number of hits in the HGTD can be processed and sent to the backend electronics by a dedicated fast pipeline system and will provide bunch by bunch luminosity information to ATLAS.

The HGTD is conceived as a modular design where the sensors can be replaced without the disassembling of the beam pipe. This has an important impact in operation as during the lifetime of the HL-LHC as the inner ring of sensors is planned to be replaced twice in order to mitigate the performance loss caused by the radiation damage. Modular design allows this operation to be performed during the end of year short shutdown periods and are planned to occur at every 1/ab step of integrated luminosity. The R\&D on LGAD sensors is planned to actively continue even after the start of HGTD operations, aiming to even harder sensors which may allow the replacement of the inner rings to occur only once.

Given the novelty of the sensors, one key aspect of high interest is the radiation hardening of the LGAD sensors in order to maximize the lifetime of the sensors in the highest radiation areas (region covering $3.5<|\eta|<4.0$). Preliminary tests point to a gain reduction of the LGADs after a fluence of about $10^{14}$~MeV n$_{\rm eq}$/cm$^{2}$. The gain can be compensated by increasing the bias in the sensor up to about 1~kV (depending on the sensor geometry and fabrication), which can further allow the sensors to operate up to $10^{15}$~MeV~neq/cm$^{2}$. This degradation occurs due to the inactivation of dopant in the gain layer, the creation of trapping centers in the crystal (which reduces the mean free path of the electrons) and in the increase of the dark current. Several techniques like the addition of Carbon, the substitution of Boron by Gallium in the doping layer (this has proved useful in space solar cells used in satellites), optimization of the gain layer (thinner but higher doped) and the controlled annealing of the devices have been proposed to increase the hardness of the sensors. This is an area where the systematic study of these effects can have a significant impact on the proposed applications of LGAD in the LHC detectors. The presence of several irradiation facilities in São Paulo provides support for this research. 

ATLAS USP group has projected and constructed a dedicated facility  for  the sensors R\&D and  has a compromise with the qualification, installation and commissioning of the detector layers of HGTD, as well as providing part of the LGAD sensor arrays needed for the innermost rings of HGTD.  The R\&D and qualification of the sensors has been done at IFUSP, partnered with the engineering groups of EPUSP and FEI. It is foreseen for the next years a contribution to the qualification and construction of the electronic boards (Peripheral Electronic Boards) that will interface the front-end electronics and the trigger and data acquisition system in this region. The installation (stave loading, module assembly and peripheral electronics integration) will be done at a CERN facility assembled exclusively for this task with the participation of members of ATLAS USP and UERJ groups, that have also participated in the  test beams at SPS (CERN) and in the construction of a small section of HGTD (demonstrator) for testing purposes at CERN. 


\subsubsection{Gas Based Detectors: From the RPC to the Future}

Gas detectors operate on the principle of ionizing a gas or a mixture of gases, which causes a controlled electrical discharge through the multiplication of electrons in an intense electric field. A significant advancement in gas detector technology is the Resistive Plate Chamber (RPC), introduced by R. Santonico and R. Cardarelli. They described it as a "\textit{dc operated particle detector (...) whose constituent elements are two parallel electrode Bakelite plates between}"~\cite{rpc_seminal}.  The key innovation of the RPC, compared to other similar gaseous detectors, is the use of two resistive plates as anode and cathode, which allows for a small localized region of dead time, resulting in excellent time resolution.

At CMS, RPCs are installed in both the barrel and endcap regions, forming a redundant system with the Drift Tubes (barrel) and Cathode Strip Chambers (endcap). According to the CMS Muon Technical Design Report ~\cite{muon_tdr}, the RPC system is composed of 423 Endcap chambers and 633 barrel chambers. Over the next four years, the CMS Muon Systems are scheduled for an upgrade. These upgrades aim to extend the pseudorapidity coverage ($\eta$) and ensure the operational conditions of the current system during the High Luminosity LHC (HL-LHC) era. For the RPC subsystem, this includes maintenance of existing chambers and the installation of new chambers in the region of $|\eta| < 1.8$ to $|\eta| < 2.4$ \cite{Pedraza-Morales_2020}.

One of the primary challenges in RPC technology is developing new gas mixtures with lower environmental impact (eco-friendly gas mixtures) while maintaining cost-effectiveness and performance that supports ionization radiation detector technologies in gaseous media for current and future HEP experiments. This includes the study of signal production performance in RPC detectors and investigating eco-friendly gas mixtures.

Another challenge of RPC technology is the development of a gas regeneration system within a closed-loop recycling circuit to mitigate the significant environmental and financial impacts associated with the use of refrigerant gases. In partnership with the Brazilian industry, specifically Recigases, we are currently developing an optimized regeneration system for MARTA-type RPCs \cite{MARTA2018, article:JINST2016}, assembled in Brazil. This project aims to conduct initial studies on the performance of the regenerated gas in signal production in these RPC detectors, thus contributing to the sustainability and efficiency of high-energy particle detection systems.

Through the INCT CERN Brasil initiative, Brazilian laboratories associated with the CMS RPC system are taking the next step in gaseous detector technology advancement and have begun upgrading aiming for the development of eco-friendly gas mixtures, gas recirculation, and a deeper understanding of how avalanches form in different gas combinations. These efforts are aligned perfectly with the goals of the DRD1 Collaboration \cite{Colaleo:2885937}, an international R\&D project dedicated to creating the next generation of gaseous detectors for particle physics experiments. Leveraging this enhanced technical environment, Brazilian researchers plan to delve even further. Their future investigations include exploring new electrode materials and the exciting potential of Micro Pattern Gaseous Detectors (MPGDs), the latest gas-based technology for high-energy particle detectors.

\subsubsection{High-Speed Electronics for Triggering and Energy Estimation}

\begin{itemize}

\item Machine learning in embedded systems for Electron and Photon Triggers

\end{itemize}

Electron and Photons in final states play an important role in LHC physics program. As leptons are key signals of electroweak interactions, it's important to record as many events containing single leptons as possible. The dihiggs production in the $HH \to b\overline{b}\gamma\gamma$ decay process is one of the key contributing channels for the overall $HH$ sensitivity, which is a key goal of the HL-LHC program. The low mass of the Higgs boson, for which precision measurements are a primary motivator for the Phase-II program, also requires these single-electron and diphoton identification thresholds to remain as low as possible \cite{ATLAS_HIGGS2021HGG}. Studies of processes involving $H\to\gamma\gamma$ will enter a new regime of precision in the HL-LHC. 


For the Phase-II upgrade \cite{CERN-LHCC-2017-019}, the ATLAS trigger systems, both hardware and software, need to be redesigned to meet expected luminosity and pileup conditions. The Trigger and Data Acquisition (TDAq) system ensures optimal data-taking conditions and selects the most interesting collision events for study. For the Phase-II, the TDAq will need to cope with a readout rate of 1~MHz, about 10 times if compared to Run-3. To achieve this, ATLAS will use a new architecture with a level-0 trigger (the first-level hardware trigger) - the Global Trigger, that will take information from a subset of information from the calorimeters and Muon Spectrometer, implementing an offline-like particle identification at hardware level. During HL-LHC running, the global trigger system will be required to handle 50~Tb/s as input and to decide within 10~$\mu$s whether each event should be recorded or discarded, allowing for more sophisticated algorithms to be run online for particle identification.

Modern machine learning techniques have shown satisfactory results in triggering during Run-2, especially implemented at the software level. The good performance of these systems has motivated the appropriate development for embedded electronics, with the aim of maintaining high efficiency in the identification of electrons and photons, and a high background noise reduction. In this context, the INCT-CERN Brazil aims to support the development of trigger systems in FPGA, given the need for scientific and technological development for the HL-LHC context. This project has the potential to combine efforts between researchers from Electronic Engineering, Physics and Computational Intelligence, allowing the integration of this knowledge in the severe tasks with the luminosity and pileup conditions foreseen for Phase-II. The activities of this development involve mainly the UFBA and UFRJ institutes, where professors and PhD students are involved in the efforts for the first proof of concepts. The NeuralRinger algorithm \cite{ATLAS_Trigger_Ringer_Run2},  Brazilian method used as standard during the end of Run-2 and the whole of Run-3 for online electron identification in the ATLAS High Level Trigger (and commissioned for photon triggers) is being implemented in simulation in the environment planned for Run-4, and is being designed for its operation in hardware. Machine learning applications also include the possibility of improving energy calibration. Proofs of cont were developed during Run-3, in which good results were obtained combining the Ringer Algorithm and Boosted Decision Trees to reduce the difference in electron energy measurements between Monte Carlo simulations and reconstructed energy. The results indicate that applications like this can be improved for hardware operation, as a contribution to the Phase-II upgrade.


\begin{itemize}

\item Energy Estimation Techniques for Hadronic Calorimeter

\end{itemize}

In the HL-LHC context, energy estimation methods play a important role on particle identification and correct event reconstruction \cite{fullana}. In high pileup environments, the energy estimation methods based on linear optimal filtering achieves a suboptimal performance, since the standard approach do not incorporate the pileup contribution to signal modeling in order to obtain the weights through the optimization process \cite{marin}. The first attempts to incorporate the pileup to signal information were based on a convolution approach, modeling the signal acquired by electronic circuits as a convolution of the signal of a giving readout window with the signals acquired in the previous bunch crossings. This approach reduce considerably the negative tail of energy estimation histograms of single calorimeter cells \cite{peralva, filho}.

Modern techniques are being proposed in order to combine machine learning approach with optimal filtering, operating in the so call free running mode. The main goal of this study is to use the performance obtained by linear methods and use a neural network optimized to correct the non-linearity produced by the pileup signal contamination \cite{InacioGoncalves:2894828}. The study is done considering a synthetic dataset in two different scenarios: $\langle \mu \rangle = 100$ and $\langle \mu \rangle = 200$. Besides evaluating the energy reconstruction histogram, the actual study aims to evaluate the impact of usage of the methods on the event actance. All the study is conducted to embed such energy estimation methods in FPGA boards. The efforts on this direction involves professors and students from UERJ and UFJF.

\subsubsection{LHCb experimental challenges}


In the last years, the LHCb detector has been upgraded to run at a five times higher instantaneous luminosity than during Run 1+2 of the LHC~\cite{LHCb-TDR-012}. 
All tracking detectors and most of the readout electronics of the subdetectors have been replaced~\cite{LHCb:2023hlw}.
This is called LHCb Upgrade I-a and there will a modest upgrade in 2025, called LHCb Upgrade I-b. 
Another major upgrade is proposed to take place in 2032~\cite{LHCb-PII-EoI}, LHCb Upgrade II.
These upgrades are motivated to provide complementary approaches of the energy and intensity frontier in the search for physics beyond the SM~\cite{LHCb-TDR-023}.  The Brazilian team is involved in the operation of three detector projects of the current experiment, especially the Upgrade I of the VELO (VErtex LOcator) subdetector based on silicon pixel detectors, and we propose to continue working on the development of these projects for the LHCb Upgrade II.

\subsubsection{The \textit{Mighty Tracker} detector for the LHCb Experiment Upgrade-II}

The LHCb Experiment underwent its first major upgrade (Upgrade I) during the second extended shutdown period of the LHC machine (LS2), which ended in April 2022. An intensive commissioning campaign lasted until the end of 2023 to improve on many world-best physics measurements of the novel technology systems added to the renovated detector. Since then, the LHCb Experiment has taken physics data from the LHC Run 3 beam with its full potential.

A smaller intervention is planned during LS3 – between 2026 and 2028 – and, finally, the implementation of the second major upgrade (Upgrade II) will take place during LS4, between 2033 and 2035. Upgrade II involves replacing a significant portion of detectors and data acquisition systems that make up the experiment. This replacement aims to address obsolescence or substantial damage caused by exposure to ionizing radiation and to handle the challenges of collecting data with even higher beam luminosity.

\subsubsection{Developments for the CMS ECAL Monitoring and Calibration system}

The CMS, as a general-purpose detector, operates with both proton beams and heavy ion beams. Around its tracking system (tracker) we have the electromagnetic calorimeter (ECAL) consisting of 75,858 lead tungstate (PbWO4) scintillating crystals, which measure and identify photons and electrons with high resolution. The performance of CMS calorimeters depends on a sophisticated monitoring and calibration system to maintain calorimeter accuracy over time by compensating for changes in detector response. 

Incorporating specific methodologies such as crystal inter-calibration using Z and W boson events decaying into electrons, continuous monitoring of the invariant mass $m_{ee}$ and $\pi^{0}$, along with pulse shape analysis, time calibration, APD noise level monitoring, and laser transparency corrections, the system expertly maintains ECAL performance. These management of these components, under the responsibility of the HI+ECAL at the Brazilian Center for Physics Research (CBPF)  project in the Monitoring and Calibration (MoCa) group, are vital to the accurate and reliable operation of ECAL, contributing significantly to the success of the physical objectives of the CMS experiment, especially in the challenging environment of the HL-LHC. 


\subsection{Contributions for computing development and Data Analyses}

\subsubsection{Next Generation Triggers}
The Next Generation Triggers (NGT) project aims to enhance the capabilities of the CMS and ATLAS experiments at the LHC by improving the trigger and data acquisition systems. The project focuses on integrating advanced computing techniques, including state-of-the-art machine learning (ML) and artificial intelligence (AI), to optimize the selection and processing of collision events in real-time. 
The project was approved by the CERN council in October 2023~\cite{CERN-council-Oct2023} and
is organised in four work packages (WP): 
\begin{itemize}
\item\emph{Infrastructure, Algorithms and Theory (WP1)}, focused on the development of software tools and methodologies for optimising event generators, experiment code and analysis algorithms, including those based on neural networks, on accelerated architectures.
\item\emph{Enhancing the ATLAS Trigger and Data Acquisition (WP2)} and \emph{Enhancing the CMS Real Time Data Processing (WP3)}, experiment-specific work packages that will redesign their data collection strategies to extend the reach of their physics programmes.
\item\emph{Education Programmes and Outreach (WP4)}, to promote exchanges across computer scientists and physics researchers, academia and industry and complement existing education programmes to train data science and AI skills for the next generation of high-energy physicists.
\end{itemize}
The SPRACE group is currently involved in WP3, with the goal of extending the reach of the CMS physics programme in view of the HL-LHC era.

\subsubsection{The ALICE O$^2$ framework}

ALICE has upgraded several of its detectors for LHC Run 3 to function in a continuous readout mode. It allows to capture Pb–Pb collisions at an interaction rate of 50 kHz without requiring a trigger. This requires processing data in real time at speeds 100 times greater than in Run 2. To address this challenge, a new computing system, called  Online-Offline (O$^2$) software framework was created. It allows parallel data processing for both the synchronous (calibration and raw data compression) and asynchronous data taking. The synchronous processing depends extensively on GPUs to deliver the computational power required to handle heavy-ion collision data in real time.

At the UNICAMP group, we have been participating in the development of the analysis framework for Hard Probes and Light Flavour physics.

\subsubsection{International Linear Collider (ILC)}

It is beyond dispute that the LHC and its plethora of physics results is a success story. And with the current machine and detectors underway, the LHC will provide even more insights to particle physics for years to come. But we also know that even a powerful machine like the LHC will have it is limitations. The center-of-mass-energy reach is bound, statistical uncertainties will not improve anymore without significantly larger data sets and systematic uncertainties are technologically constrained and eventually capped by limited statistics.

The HEP community is in agreement that only a new accelerator and detector will help to push the limits of particle physics research. And although the LHC will be provided data for at least another decade, it is crucial to study, coordinate and plan future accelerators. With the International Linear Collider (ILC) community, we have one of the strongest groups that is working on these tasks. The ILC efforts circling around an electron-positron collider with a center-of-mass energy of the order of $\mathcal{O}$(500~GeV). Japan is currently the favored host of such a machine and its associated laboratory. It is foreseen that two complimentary detectors will record collisions in a so-called push-pull configuration. The project, its physics potential and technical details are described in~\cite{Behnke:2013xla, Baer:2013cma, Adolphsen:2013jya, Adolphsen:2013kya, Behnke:2013lya}.

\begin{figure}[h]
\centering
\includegraphics[width=8cm]{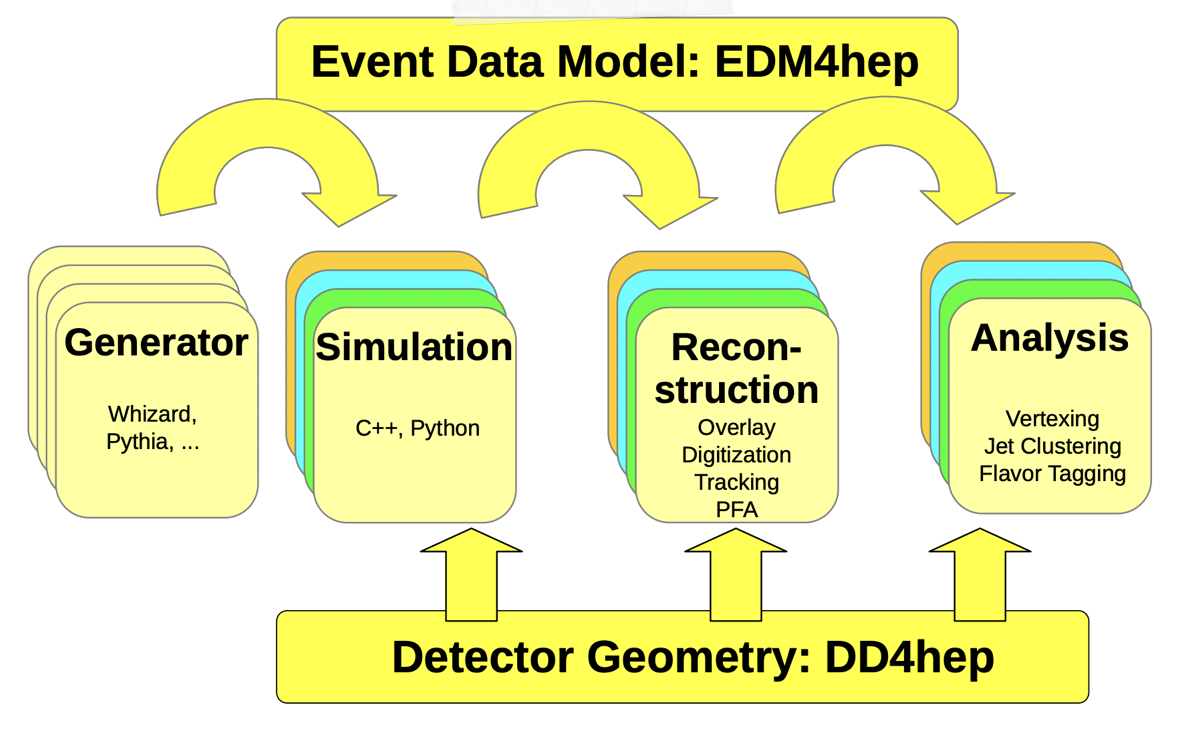}
\caption{Illustration of the key4hep software stack.}
\label{fig:key4hep}
\end{figure}

Key4hep~\cite{Key4hep:2023nmr}, a turnkey software for future colliders, is currently under development at DESY and CERN. It is a software stack that connects and extends individual packages towards a complete data processing framework for detector studies. It is made up out of: (i) fast and full simulation, (ii) reconstruction software, and (iii) analysis software. The components of key4hep are: (i) event data model: EDM4hep, (ii) geometry information: DD4hep, (iii) analysis framework: Gaudi, and (iv) packaging and deployment: Spack.
Figure~\ref{fig:key4hep} illustrates the Key4hep software.

During the last year, we started writing a software package using key4hep that analysis MC data with the goal of identifying the production of a Higgs boson and its subsequent decay into four neutrinos ($H\to ZZ^\star\to \nu\bar{\nu}\nu\bar{\nu}$). These so-called invisible Higgs decays are very rare in the SM of Particle Physics (branching ratio$\sim\mathcal{O}(0.1\%)$). LHC can only set upper limits on this decay channel~\cite{ATLAS:2022yvh, CMS:2020ulv}. 
Studying this kind of decays using key4hep serves two purposes:
\begin{enumerate}
    \item Study the sensitivity of the various detector proposals. Key4hep allows to run the same analysis code on different detector simulations. In the past, similar analyses were done that lead to contradictory results. This analysis will resolve these contradictions.
    \item This is the first time key4hep is used under realistic conditions. The analysis serves as a test for the software stack and tries to identify problems and shortcomings.
\end{enumerate}

\subsubsection{Future Circular Collider (FCC)}

The Future Circular Collider (FCC) represents an international collaboration aimed at pushing the boundaries of particle physics and exploring new frontiers of scientific knowledge. The project involves the participation of research institutions from around the world, united by the common goal of developing and implementing the next generation of particle colliders. The first stage, FCC-ee, is an electron-positron ($e^{+}e^{-}$) collider designed to operate at center-of-mass energies ranging from below the Z boson pole (90~GeV) to beyond the top-quark pair-production threshold (365~GeV), followed by a hadron-hadron collider (FCC-hh) that would enhance the direct discovery potential for new particles.

The Brazilian participation in the FCC will provide training and development for a new generation of scientists and engineers, preparing them to face the technological and scientific challenges of the future. CBPF has been an official and active participant in the FCC collaboration since 2016, being the first Brazilian institution to officially collaborate with the FCC project. CBPF has been involved in feasibility studies, the development of data analysis algorithms and detector simulations, fine-tuning of Monte Carlo simulators, and training of future collaborators, playing a crucial role in the project's progress with analyses included in all its activity reports~\cite{FCC:2018byv, FCC:2018evy, FCC:2018vvp, FCC:midterm, RebelloTeles:2023uig, dEnterria:2019jty}.

In 2023 the FCC collaboration welcomed a new Brazilian participant, the Federal University of Rio Grande do Norte (UFRN). The research group at UFRN focuses on new physics studies related to dark matter production~\cite{LHeC:2020van,Huang:2022ceu}, new heavy resonances featuring flavor violation, demonstrating that the FCC can be competitive with known restrictive flavor physics probes \cite{CarcamoHernandez:2022fvl} and projecting FCC limits on potential new physics interpretations \cite{deJesus:2023som} motivated by the recent measurement of the muon anomalous magnetic moment by the Muon g-2 experiment at Fermilab.




\section{Outreach initiatives for fostering new researchers}

Attracting new students to the Physics courses has been a challenge worldwide, and pursing novel ways to advertise and communicate about Science with young students is a main goal for the scientific community. The INCT CERN-Brazil has a dedicated working group to coordinate actions in Brazil to take the LHC Physics everywhere. Also, training young undergraduate students is an important task to allow new researchers to start their projects within the CERN Collaboration. In INCT CERN-Brazil a Data Analysis School is being organized to take place in November/2024 to prepare new researchers in using modern tools for data analysis.

\subsection{Northern Brazil: UFRN (Natal)}
Our community has been actively organizing the International Masterclass in Particle Physics across the country. We have many examples of Ph.D. students who have been fished in Masterclass events devoted to high school students. We also plan to expand the Masterclass program to high school teachers. Moreover, we have established a set of actions to attract new members and raise interest in collider physics. In particular, our community will host important events in the upcoming years: (i) In February 2024, the International Institute of Physics-UFRN hosted a Brazilian School on Collider Physics. In September, the Institute will also promote an international school on collider physics; (ii) In 2025 the International Institute of Physics-UFRN expects to host an International School in accelerator physics; (iii) In 2026, the Brazilian community is proud to be host the International Conference on High Energy Physics (ICHEP2026). The conference will also take place in Natal, the city which hosts the International Institute of Physics-UFRN; and (iv) In 2026 we also expect to host in Natal the IDPASC International School in High Energy Physics shortly after ICHEP2026.

Therefore, it is rather clear that our community is engaged in forming new students and creating new opportunities.





\subsection{Central Brazil: CBPF (Rio de Janeiro) and SPRACE (São Paulo)}


CMS collaboration is a pioneer in the release of LHC data for public use. In 2020 and 2021, the collaboration provided workshops, where theorists, phenomenologists, and other interested researchers were able to learn more about how to interact and analyze these vast data sets. This is the first contact of undergraduate students and theoretical postdoctoral with the \textit{modus operandi} of an experimental analysis. The HI+ECAL@CBPF project has been coordinating the Master Class Hands-On in Particle Physics at CBPF, in collaboration with the Fluminense Federal University (UFF) and SPRACE in São Paulo (the pioneer of Master Class in Brazil), promoting the dissemination of scientific knowledge. Our wide outreach actions and training of human resources in CBPF, UFF and SPRACE are also focused on encouraging female participation in science, supported by projects submitted to CNPq. Our main goal is to broaden the influence of these activities among young people, with a special focus on the inclusion of girls, who are still underrepresented in this area of expertise. This initiative represents an important step towards the diversification and strengthening of the scientific and technological future.




\subsection{Southern Brazil: UFRGS (Porto Alegre)}

The IPPOG Masterclass has been an efficient activity to allow young student from High Schools to engage on the topics being investigated at the LHC\footnote{\url{https://ufrgs.br/cms/masterclass}}. UFRGS has a partnership with the Instituto Federal in Osório, Rio Grande do Sul, to organize the Masterclass every year, with 2 editions already realized. Around 60 students have participated in each event to learn, study, and discuss the search of particle in the CMS experiment. Besides, UFRGS organized every year an event named \textit{Open Doors} to allow all the local community to visit its laboratories and learn more about the investigations carried out. Since 2019, the LHC experiments prepared presentations and experiment demonstrations to introduce Particle Physics to the community, with very good feedback received.

\bibliography{refs}
\bibliographystyle{JHEP}
\end{document}